\input{epsf}
\epsfverbosetrue
\documentstyle[prl,aps,multicol,epsf]{revtex}
\renewcommand{\narrowtext}{\begin{multicols}{2} \global\columnwidth20.5pc}
\renewcommand{\widetext}{\end{multicols} \global\columnwidth42.5pc}
\begin{document}
\draft

\title{``Cosmological'' scenario for $A-B$ phase transition in superfluid
$^3$He.}
\author{Yu. M. Bunkov,  O. D. Timofeevskaya
\cite{bTimof}.}
\address{ Centre de Recherches sur les
Tr\`es Basses Temp\'eratures - CNRS,
%\\Laboratoire associ\'e \`a l'Universit\'e Joseph Fourier,
BP166, 38042, Grenoble, France.}

\date{\today} \maketitle

\begin{abstract}
 At a very rapid superfluid transition in $^3$He, follows after a reaction
with single neutron, the creation of
topological defects (vortices) has recently been demonstrated in accordance
with the
Kibble-Zurek scenario for the cosmological analogue. We discuss here the
extension of
the Kibble-Zurek scenario to the case when alternative symmetries may be
broken and
different states nucleated independently. We have calculated the nucleation
probability of
the various states of superfluid $^3$He during a superfluid transition. Our
results can
explain the transition from supercooled $A$ phase to the $B$ phase,
triggered by nuclear
reaction. The new scenario is an alternative to the well-known ``baked
Alaska'' scenario.
\end{abstract}

\narrowtext

 Superfluid $^3$He has an order parameter which describe the simultaneous
spin, orbital
and gauge symmetries which are broken at the superfluid transition. This
transition can
be regarded as the closest condensed matter analogy to the cosmological
grand unification
transition. This analogy have been utilised in the experimental test of the
Kibble
cosmological mechanism of cosmic strings creation. According to this mechanism
\cite{1}, at the transition separate regions of the Universe are
independently nucleated
with a random orientation of the gauge field in each region. The size of
these initial
regions (domains) depends strongly on the rapidity with which the
transition is traversed.
According to Zurek \cite{2} the fundamental distance between the
independently-created
coherent domains (in the language of \cite{2} the distance between the
ensuing vortices
$Z$) is of the order of $Z=\xi_0(\tau_Q/\tau_0)^{1/4}$, where $\xi_0$ is
the zero
temperature coherence, $\tau_0 = (\xi_0/v_F)$ is the characteristic time
constant of the
superfluid, and $\tau_Q$ is the characteristic time for cooling through the
phase
transition. As the domains grow and make contact with their neighbours, the
resulting
gauge field cannot be uniform. The subsequent order-parameter ``glass''
forces a
distribution of topological defects leading to a tangle of quantized vortex
lines. The first
quantitative tests of defect creation during a gauge symmetry
transformation have been
recently performed in superfluid $^3$He.

The superfluid $^3$He (at very low temperature in the Grenoble experiment
\cite{3} and
at a relatively high temperature in the Helsinki experiment \cite{4}) was
heated locally by
neutron irradiation via the nuclear reaction:
\centerline{$^3$He + n = $^3$H$^-$ + p$^+$ + 764 keV} The energy released by a
neutron reaction heats a small region of the liquid $^3$He (about 30$\mu
m$) into the
normal state. This region recools rapidly through the superfluid transition
owing to the
rapid outflow of quasiparticles into the surrounding superfluid. For the
experimental
conditions of both experiments it has been proposed that quasiparticles
from the heated
region disperse outwards, meaning that the hot bubble is cooled rapidly
from its sides
and that the cooling rate is so fast that the order parameter of the
surrounding superfluid
$^3$He cannot follow the changing temperature front fast enough (see
\cite{5} for
theoretical details). Consequently internal regions of the hot volume transit into the
superfluid phase independently in accordance with the Zurek cosmological
scenario. The
experimental results of both experiments justify this assumption. In the
Grenoble
experiment the excess number of quasiparticles created by the reaction has
been counted
and it was found that a significant fraction of the energy released by the
reaction does
not appear in the quasiparticles thermal reservoir. This energy deficit
agrees well in
magnitude with the energy expected to be trapped as topological defects (in
this case
vortices) as calculated from Zurek's scenario for the Kibble mechanism.

Under the relatively high temperature conditions of the Helsinki experiment
any vortices
created by the neutron reaction would be rapidly destroyed via interaction
with the
quasiparticle gas. However, in the ROTA project rotating cryostat there is
an added bias
field, that of rotation. This field can extract a few vortex rings from the
bubble which
then grow to the dimensions of the cell. After the process the number of
vortices can be
measured directly by NMR. The number of extracted vortices corresponds well
to that
calculated from the Zurek scenario.

Our knowledge of superfluid $^3$He is much better than our knowledge of the
Universe. In the case of superfluid $^3$He we not only know the symmetries
broken
during the superfluid transition but we also know the Ginzburg-Landau
potential exactly
and we can calculate quantitatively the dynamics of the order parameter
during the
transition. There are two different stable phases of $^3$He, the $A$ and
$B$ phases
which correspond to different broken symmetries. The energy difference
between these
two states is relatively small. Let us say that it is negligible on the
timescale of the
transition! This means that regions which independently enter the
superfluid state, should
not only have different orientation of the order parameter but may also
correspond to states with different symmetries
\cite{6}. It is this complication of the Kibble-Zurek scenario which we
considered in the
calculations below. Ironically, a very similar situation may be relevant to
the Universe,
where in addition to the creation of the $SU(3) \times SU(2)\times U(1)$
state, other
states may also be created, in particular, $SU(4) \times U(1)$ state
\cite{7}. The first
state, we believe, corresponds to the energy minimum of our Universe,
whereas the
second state has much higher creation probability owing to its higher
symmetry. This is
exactly the situation in superfluid $^3$He where the $B$ phase has the
lower energy,
except in the case of the strong interaction correction for high pressure
and temperature.

The rotational and gauge symmetries of $^3$He are usually represented by a
3$\times$3
matrix of complex numbers $A_{ai}$ which is known as the order parameter. Above
the transition all the elements of the matrix have zero values
(representing full symmetry).
Below the transition, some of these quantities become non-zero. The
symmetry of the
order parameter after the transition corresponds to the manifold of
symmetries which
remain unbroken. In the case of superfluid $^3$He there are 13
possibilities (13 states)
corresponding to the various symmetries of the order parameter \cite{8}.
The free energy
of these states can be expressed in the framework of the phenomenological
theory of
Ginzburg and Landau by:

$F= -\alpha A^*_{ai} A_{ai} + \beta_1 A^*_{ai} A^*_{ai} A_{bj}
A_{bj}+ \beta_2 A^*_{ai} A_{ai} A^*_{bj} A_{bj} + \beta_3
A^*_{ai}
A^*_{bi} A_{aj} A_{bj} + \beta_4 A^*_{ai} A_{bi} A^*_{bj}
A_{aj} +
\beta_5 A^*_{ai} A_{bi} A_{bj} A^*_{aj} $

\noindent where $\alpha = \alpha_0 (1 - T/T_c)$ changes sign at the transition
temperature $T_c$, and the quantities $\beta_i$ are functions of pressure
(and also of
temperature through the so-called ``strong correction'') and depend on the
details of the
microscopic interaction.

The different possible symmetries of the order parameter correspond to
local minima and
saddle points in this 18-dimensional energy surface.  In superfluid
$^3$He we know there are two stable states, the $A$ and $B$ phases.  The
energy balance between the
$A$ and $B$ phases is determined by the relation between the parameters
$\beta_i$. At
zero pressure, the $B$ phase corresponds to the absolute minimum, while at
pressures
above 20 bar there is a region of temperature where the A phase becomes the
preferred
state.

These two states have different order parameter symmetries. In the $B$
phase, relative
spin-orbit symmetry $SO(3)_{S+L}$ remains unbroken (such that $A_{ai}$
resembles
a rotation matrix). In the $A$ phase (the ``axial'' state) the symmetry of
the spin system is reduced to a
gauge symmetry, which couples to the orbital motion to yield a combined
symmetry of
the orbital rotation and gauge fields $U_S \times U_{L+G}$.

According to Zurek scenario, regions on a distance scale of $Z$ undergo the
superfluid transition separately. We can consider these regions as
independent elementary
samples of $^3$He. (Later we shall analyse the influence of the gradient
energy between
the different regions.) We have numerically modelled the creation of the
superfluid
phases in a single region during a rapid superfluid transition. We applied
a small random
perturbation to the $A_{ai}$ matrix at $T=T_c$. Then we have reduced the
temperature with some
velocity and have calculated the development of order parameter during this
``downhill''
process. For this we have applied the time dependent Ginzburg-Landau
equation in the
form:

$- \tau {\partial \over \partial t} A_{a,i} + {T_c - T(t) \over T_c}
A_{a,i} - ( \beta_1 A^*_{ai} A_{bj} A_{bj}+ \beta_2 A_{ai}
A^*_{bj}
A_{bj}\\
+ \beta_3 A^*_{bi} A_{aj} A_{bj}
+ \beta_4 A_{bi} A^*_{bj} A_{aj}
+ \beta_5 A_{bi} A_{bj} A^*_{aj}) = 0$

We have monitored both the symmetry of order parameter and the energy
during this
time-evolution. We have found that both the $A$ and $B$ phases (as well as the
axi-planar state at 0 bar, see below) can develop. The final state depends
on the starting
orientation of the order parameter and the profile of the 18-dimensions
potential surface.
Other metastable states may develop transiently after the application of an
initial
perturbation which has the exact symmetry of these states. However the
trajectory of
$A_{ai}$ in these cases is unstable and any small perturbation away from
the final
symmetry leads to the more stable $A$ or $B$ states.

\begin{figure}[htb]
\centerline{\epsfxsize=8 cm \epsfbox{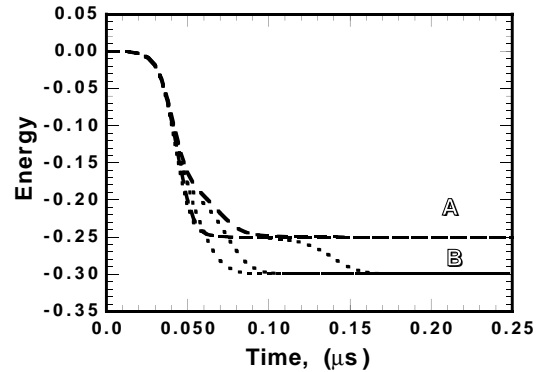}}
\bigskip
\caption{The time evolution of the free energy density during a superfluid
phase transition after a small random perturbations.
The temperature was reduced from $T=T_c$ to $T=0$ in a time of
$10^{-8}$s. Dashed lines correspond to transitions resulting in A phase, while
dotted lines end in B phase.}
\label{fig1}
\end{figure}

It is important to note that, although according to Zurek the cooling rate
determines the
dimensions of the independent regions, the trajectory of the order parameter
for a single coherent region is rate independent and is determined only by
the profile of
the G-L potential.
At zero pressure, when we only have the weak interaction where
 $\beta_i = (-1, 2,2,2, -2)$, the $B$ phase corresponds to the absolute
energy minimum. In our computer simulation
we find that, even under these conditions, nucleation of the $A$ phase has
a high
probability. In quantitative terms we find the probability of $B$ phase
creation to be
$54\% \pm 1 \%$, while that of the $A$ phase creation is $46 \%$. It is
difficult to
visualise the trajectory of the order parameter in 18 dimensional space,
but we can
monitor the G-L energy during the transition. Fig.1 shows typical
trajectories of the
superfluid $^3$He free energy after rapid cooling. In some cases the trajectory
approaches regions of saddle points on the energy surface.
The behavior here is clarified by reducing the rate of energy change.

For to study the influence of gradient energy on the development of the order
parameter we have considered a one-dimensional spatial sample of Zurek length $Z$
divided into 100 points. We have chosen $Z$ to agree with the Grenoble
experiment at
zero bar (about 8$\xi_0$). Two different perturbations have been applied,
one for the
first 50 points and the other for the second 50 points. The development of
the $A_{ai}$
matrix during the
``downhill'' process has been calculated at each point, taking into account
the gradient
energy. The results of these calculations, when a perturbation with A-phase
symmetry is
applied to one side, and with a $B$-phase symmetry on the other, are shown
in Fig.2.
We have
found that the boundary between the two different states remains almost
stationary during the main part of the ``downhill'' process. Towards the end of
this process the boundary begins to move in the energetically favourable
direction. This
result looks very natural, since the boundary replacement is determined by
the energy
difference, and the time dependence of the energy is very similar for the
two different
symmetries at the
beginning of the ``downhill'' process (see Fig.1).

\begin{figure}[bt]
\centerline{\epsfxsize=7 cm \epsfbox{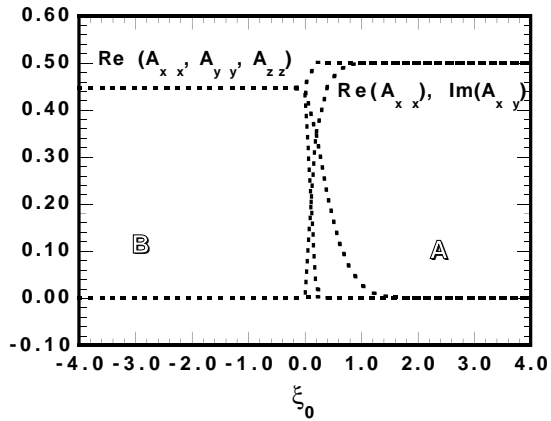}}
\bigskip
\caption{The nonzero terms of the order-parameter evolution during a
superfluid phase
transition. At time zero a small perturbation was applied (with $B$-phase
symmetry for
the left hand side and $A$-phase symmetry for the right hand side). }
\label{fig2}
\end{figure}

As was discussed by Volovik \cite{9}, the $A$ phase at 0 bar has an
additional hidden symmetry. It is correspond to the degenerate manifold of
states between ``axial'' state and ``planar'' state. The planar state is
corresponds to saddle point, it does not separated by potential barrier
from  the $B$ state. Nevertheless the domain boundary forms even between
``planar'' and $B$ states, but then moved relatively  fast. With pressure
the degeneracy removed in favour of ``axial'' state.

The $\beta_i $ parameters depend on the pressure and temperature. There are
a number of
theories which suggest somewhat different dependencies of these parameters
on pressure
at T$_c$. We have used the parameters, calculated by Sauls and Serene
\cite{10}.
In Fig.3 we show the probability of
$A$ phase nucleation as a function of pressure along with the energy
balance between the
$A$ and $B$ phases. It is important to notice that the probability of $A$
state nucleation
may become greater than 50\%{} even in the region where the $B$ phase is stable.

The
temperature dependence of $\beta_i $ has not been much investigated
theoretically.
Qualitatively, we expect that they should change in the same way as the A-B
equilibrium
line changes on the $^3$He phase diagram. This would imply that with
cooling the
strong interaction correction should decrease very rapidly. The temperature
dependence
of $\beta_i$ parameters is in fact very important for our scenario, because
under the
non-equilibrium conditions of the fast phase transition the temperature
changes faster than
the order parameter.

\begin{figure}[htb]
\centerline{\epsfxsize=8 cm \epsfbox{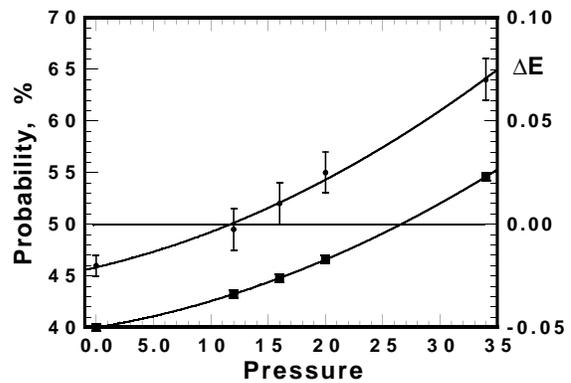}}
\bigskip
\caption{The probability of $A$ state nucleation as function of pressure
for temperature
near $T_c$, and the difference of energy between $A$ and $B$ states.}
\label{fig3}
\end{figure}

All experimentalists who work with superfluid $^3$He have noticed the crucial
asymmetry of the $A-B$ transition. If one is cooling $^3$He at a pressure
above 20 bar,
the $A$ phase may survive as a supercooled metastable state far below the
equilibrium
$A-B$ transition line. On the other hand, on warming it is difficult to get
superheated
$B$ phase. In \cite{11} it was shown that a transition from $A$ to $B$
phase will
always occur at some critical temperature. The pressure dependence of this
threshold temperature is parallel to the equilibrium A-B transition line.
It crosses the $T_c$ temperature line at about 15 bar. This
corresponds well to the situation in our calculations where the probability
of $B$
nucleation exceeds that of $A$ phase nucleation.

This observation may supply the critical jigsaw piece of information for the long-running
puzzle of the $A-B$ transition in superfluid $^3$He. As proposed by Leggett and
demonstrated in the Stanford experiments (see review \cite {12}) cosmic
rays can trigger
the transition from supercooled $A$ phase to $B$ phase. In the well-known
``baked
Alaska'' scenario, proposed by Leggett \cite {12}, it is assumed that a
shell of normal liquid
expands from the reaction site. After the shell has passed the temperature
inside falls
below $T_c$ and a new state nucleates. The presence of the expanding normal
shell is
needed to isolate the nucleation of a new phase from any influence of the
surrounding
$A$ phase.
From our point of view, this is a rather artificial suggestion. It is
likely that the cosmic event creates very energetic quasiparticles. These
energetic
quasiparticles travel out from the site of the event and create many new
low energy
quasiparticles on thermalization. It is important to point out that the low
energy
quasiparticles do not maintain the direction of the primary energetic ones.
That is why it
is likely that the quasiparticles remain inside the hot bubble and expand
by the usual
diffusion process.

However, in framework of the cosmological Kibble-Zurek approach we do not
need such
a normal shell to protect the interior of the hot bubble from the influence
of the outside
state. The diffusion cooling runs so fast that many seeds of $A$ and $B$
phase are
nucleated independently. The ``backed Alaska'' process, if it occurs, would
lead to an
even larger number of such domains. The subsequent development of the structure
depends first on the relative densities of the two phases and secondly on
the energy
balance between them and on the domain boundary surface energy. If one
state has a
significantly higher probability of nucleation than the other, then
percolation occurs and
the more probable phase grows at the expense of the less probable to reduce
the surface
energy. That is the reason for the asymmetry in the $A-B$ transition. On
cooling, the
$A$ phase can be supercooled because after a cosmic rays events
the seeds of $B$ state do not survive in conditions
where the $A$ phase has a higher nucleation probability.
For to pass throw transition, the seeds of $B$ states should form a
cluster of  critical dimensions, which is possible only when $B$ state
nucleation probability is near to 50\%.

In the case where there is a possibility of nucleating two distinct phases,
then owing to
the eventual suppression of one phase, the distance between the subsequent
vortices
which remain from the order-parameter ``glass' will be larger than that
implied by the
straightforward Zurek scenario. A simple argument suggests that the
separation increases
by of order $Q^{-0.5}$, where $Q$ is the probability of nucleation of the
surviving
state. This correction makes the calculated distance between vortices
closer to that
observed in the Grenoble \cite{3} experiment. Furthermore, the influence of the
proximity of the $A$ phase on the density of vortices created has recently been
demonstrated \cite{13}

Having considered superfluid $^3$He we should look more carefully at similar
possibilities for the early Universe. In other words, the vacuum of the
Universe after a
grand unification transition may also have had metastable states with different
symmetries. For example vacua with symmetries (SU(4)$\times$U(1)) and
(SU(3)$\times$SU(2)$\times$U(1)) might have been able to coexist in the
early Universe
in separate domains. The spatial scale of these domains should be of the
order of the
parameter $Z$ in Zurek's scenario. The transition of the metastable phase
to the stable
might have given rise to temperature and density inhomogeneities which may have
influenced the Universe inhomogeneity observed at present.

 We are grateful to A.~J.~Gill, H.~Godfrin, S.~N.~Fisher,
G.~R.~Pickett and G.~E.~Volovik for many stimulating discussions.

%\begin{thebibliography}{95}

\widetext
\end{document}